\documentclass[%
 reprint,
 amsmath,amssymb,
 aps,
]{revtex4-2}

\usepackage[dvipdfmx]{graphicx}
\usepackage{graphicx}
\usepackage{dcolumn}
\usepackage{bm}
\usepackage{braket}
\usepackage{array}
\usepackage{amsmath}
\usepackage{color}

\begin{document}

\preprint{APS/123-QED}

\title{Quantum Algorithm for Shortest Vector Problems with Folded Spectrum Method}

\author{Kota Mizuno}
\email{al20013@shibaura-it.ac.jp}
\author{Shohei Watabe}
\email{watabe@shibaura-it.ac.jp}

\affiliation{Graduate School of Engineering and Science, Shibaura Institute of Technology, Toyosu, Tokyo 135-8548, Japan}

\begin{abstract}
  Quantum annealing has been recently studied to solve the shortest vector problem (SVP), where the norm of a lattice point vector is mapped to the problem Hamiltonian with the qudit encoding, Hamming-weight encoding, or binary encoding, and the problem to find the shortest vector is mapped to a problem to find a non-trivial first excited state. 
  We here propose an alternative encoding and alternative quantum algorithm to solve the SVP: the one-hot encoding and the quantum imaginary-time algorithm with the folded spectrum (FS) method. 
  We demonstrate that our approach is applicable to find the shortest vector with a variational quantum algorithm.
  The application of the FS method to the quantum annealing and simulated annealing is also discussed to solve the SVP. 
  Our study shows wide potential applicability of the SVP in quantum computing frameworks. 
\end{abstract}

\maketitle


\section{Introduction}

Post-quantum cryptography (PQC) is one of the important research area not only in the field of engineering and science but also in our modern society. 
This is because the RSA (Rivest--Shamir--Adleman) cryptosystem~\cite{10.1145/359340.359342}, the security of which relies on the difficulty of the factoring problem with two large prime numbers, will be efficiently attacked by the fault-tolerant quantum computer with the Shor's algorithm~\cite{Shor99}. 
For the PQC standardization, the lattice-based cryptography is one of the main candidate algorithms~\cite{915326}. 
In particular, three of four candidates selected for standardization, CRYSTALS-KYBER, CRYSTALS-Dilithium and FALCON, are lattice-based~\cite{915326}. 

The shortest vector problem (SVP) is one of the standard lattice cryptography. 
This problem is to find the non-trivial non-zero vector whose norm is minimum in lattice points, 
which is known to be NP-hard~\cite{10.1145/237814.237838,10.1145/276698.276705}. 
Attacking the PQC with quantum algorithm is an important issue, and 
the Grover's algorithm is applied to the SVP to find the shortest vector~\cite{Laarhoven2013}, 
whereas the computational complexity is exponential but faster than the classical algorithm. 

Alternative quantum method to attack the SVP has been discussed with the used of the quantum annealing~\cite{Joseph20,David21,Ura22}. 
The square of the norm of the lattice points are embedded in the problem Hamiltonian through the various kind of encoding, 
such as the qudit encoding~\cite{Joseph20}, the Hamming-weight encoding~\cite{David21}, and the binary encoding~\cite{David21}. 
While the conventional quantum annealing~\cite{Kadowaki98} is employed to find a non-trivial ground state, 
the ground state of the SVP Hamiltonian corresponds to the trivial zero-vector, where we need to find a non-trivial shortest vector state, 
which is the non-trivial first excited state. 
In the earlier study~\cite{Joseph20,David21}, to find the first-excited state, the non-adiabatic quantum annealing is applied, 
where the so-called ``Goldilocks zone" has been extracted~\cite{David21}, where the finding probability of the first excited state is higher than those of the other states. 
However, it is hard to optimize the annealing scheduling to produce the ``Goldilocks zone", and to reproduce the ideal zone expected in the numerical simulation is quite difficult by using the D-wave 2000Q quantum annealer~\cite{8728085,David21}. 
The performance also depends on the encoding we employ~\cite{David21}. 
The excited state quantum annealing is also proposed to solve the SVP~\cite{Ura22}, 
where preparing a first excited state as an initial state is in contrast to the conventional quantum annealing. 
In this excited state quantum annealing, keeping the first excited state during the adiabatic annealing will be a difficult task in real quantum devices. 

We here propose alternative methods to solve the SVP in quantum computers. 
Since the encoding is crucial for the performance, 
we first introduce an alternative choice of the encoding, one-hot encoding, for the SVP. 
To find the first excited state, we propose to use the quantum imaginary-time evolution (QITE) with the folded spectrum (FS) method. 
The QITE is implemented in the noisy intermediate-scale quantum (NISQ) algorithm, and the excited state search method is also proposed in NISQ algorithms~\cite{Tsuchimochi23,Nakanishi2019,Tazi2024}. 
We demonstrate that our approach is applicable to find the shortest vector with a variational quantum algorithm. 
The FS method requires to find an appropriate value of the parameter, at which the spectrum in the original Hamiltonian is folded. 
We propose the search and bound to find an appropriate value of the parameter in the SVP.  
We also dicuss the use of the quantum annealing and the simulated annealing with the FS method. 
Our study demonstrates that the SVP mapped in the Hamiltonian can be attacked not only by the quantum annealing but also by a variety of quantum algorithms, 
which pushes the further study of the PQC with the quantum computers.

\section{Shortest Vector Problem}

Let $\mathcal{L}$ be a set of all integer linear combinations of $n$ vectors $\bm{b}_1, \ldots, \bm{b}_n$ in the vector space $\mathbb{R}^m$, defined as 
\begin{align}
  \mathcal{L}(\bm{b}_1,\ldots,\bm{b}_n) \equiv
  \left\{
  \sum_{i=1}^n x_i \bm{b}_i \in \mathbb{R}^m : x_i \in \mathbb{Z}
  \right\}. 
\end{align}
If $\bm{b}_1, \ldots, \bm{b}_n \in
  \mathbb{R}^m$ are linearly independent, 
the discrete set $L\equiv \mathcal{L}(\bm{b}_1,\ldots,\bm{b}_n)$ is a lattice in the dimension $n$, where a set $\{ {\bm{b}_1, \ldots, \bm{b}_n} \}$ is a lattice basis. 
A lattice point (or lattice vector) $\bm{v}$, which is an element of the lattice, can be expressed as 
\begin{align}
  \bm{v} =
  \bm{x} \cdot \bm{B} =
  x_1 \bm{b}_1 + \cdots + x_n \bm{b}_n \in \mathcal{L}(\bm{B}),
\end{align}
where $\bm{x}=(x_1,\ldots,x_n)$ is a set of integer, i.e., $x_i \in \mathbb Z$, 
and $\bm{B} \equiv (\bm{b}_1 , \cdots , \bm{b}_n)$ is the lattice basis matrix, which provides the the lattice $\mathcal{L}(\bm{b}_1,\ldots,\bm{b}_n)=\mathcal{L}(\bm{B})$.

\begin{figure}[tbp]
  \includegraphics[keepaspectratio, width=60mm]{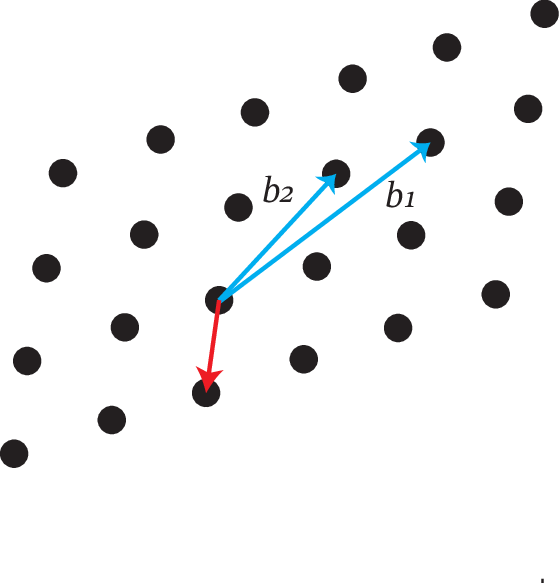}
  \caption{Schematics of the shortest vector problem (SVP) in the two-dimensional lattice. 
  Dots are the lattice point vectors. The basis vectors $\bm{b}_{1,2}$ are blue and the shortest vector is red. }
    \label{SVP}
\end{figure}

The shortest vector problem (SVP) is to find a closest non-zero vector lattice point $\bm{v}$ to the origin among the lattice point in $L$, 
given as (Fig.~\ref{SVP}) 
\begin{align}
    \lambda_1 (L) = \min ( \| \bm{v} \| : \bm{v} \in L \backslash \{ \bm{0} \}). 
\end{align}
In the optimization problem, the objective of the SVP is to find a set of integer $\bm{x}$ that minimizes the norm of non-zero vectors $\bm{v}$, where the square of the norm of $\bm{v}$ is given by 
\begin{align}
  \| \bm{v} \|^2 & = \sum_{i,j=1}^n x_ix_j \langle\bm{b}_i,\bm{b}_j \rangle \\
                 & = \sum_{i,j=1}^n x_ix_jG_{ij}. 
\end{align}
Here, $G_{ij} = \langle \bm{b}_i , \bm{b}_j \rangle$ represents the $(i,j)$-element of the Gram matrix $\bm{G}$ for a lattice basis, given by 
\begin{align}
  \bm{G} \equiv \bm{B}^T\bm{B} =
  \begin{pmatrix}
    \langle \bm{b}_1 , \bm{b}_1 \rangle & \cdots & \langle \bm{b}_1 , \bm{b}_n \rangle \\
    \vdots                              & \ddots & \vdots                              \\
    \langle \bm{b}_n,\bm{b}_1 \rangle   & \cdots & \langle \bm{b}_n,\bm{b}_n \rangle
  \end{pmatrix}. 
\end{align}

\section{Encoding of SVP to Quantum Algorithms}

We discuss the mapping of the SVP to quantum systems. 
The Euclidean norm of a vector is mapped to the energy of the generalized Bose--Hubbard Hamiltonian for an ultracold bosonic gas in an optical lattice~\cite{Joseph20}. The local occupation number $n_i$ in a site $i$ is mapped to the integer $x_i$, which plays a qudit in the quantum system. 
The qudit encoding is expanded to the qubit enconding, where two types of the mapping, the Hamming-weight encoding and binary encoding, are proposed~\cite{David21}. 
In these encodings, the Hamiltonian mapped from the norm of a vector can be given by 
\begin{align}
  \hat{H}_{\rm p} = J \sum_{i,j=1}^n G_{ij} \hat{Q}^{(i)}\hat{Q}^{(j)}, 
\end{align}
where a diagonal operator $\hat Q^{(i)}$ is mapped from the integer $x_i$, the form of which depends on the type of the encoding. 
The constant $J$ is renormalized as a unit of the energy. 
The ground state of $\hat{H}_p$ corresponds to the zero vector, i.e., $\|\bm{v}\|^2=0$, so the solution to the SVP should be the first
excited state of $\hat{H}_p$. 
In this paper, we set the coefficient to $J=1$. 

The Hamming-weight encoding represents the integer $x_i$ as the sum of the eigenvalues of a Pauli-$Z$ operator $\hat \sigma_z^{(p,i)}$~\cite{David21,Ura22}, given by 
\begin{align}
\hat Q^{(i)} = \sum\limits_{p=1}^{2k} \frac{\hat \sigma_z^{(p,i)}}{2}, 
\end{align} 
which gives an integer $x_i \in [-k, k]$. 

The binary encoding gives the integer $x_i$ of the decimal number as the binary representation, where 
the operator is given by 
\begin{align}
\hat Q^{(i)} = \sum\limits_{p=1}^{k} 2^p \hat q^{(p,i)} - 2^k \hat I, 
\end{align} 
where $\hat I$ is the identity matrix and $\hat q^{(p,i)}$ is an operator for $(p,i)$ defind by 
\begin{align}
\hat q^{(p,i)} = \frac{ \hat \sigma_z^{(p,i)} + \hat \sigma_0^{(p,i)} }{2}. 
\label{qubo}
\end{align} 
Eigenvalues of the operator $\hat q^{(p,i)}$ is $+1$ and $0$, 
and this encoding provides an integer $x_i \in [ -2^k, 2^k - 1]$. 
The diagonal operator $\hat Q^{(i)}$ can be reduced into 
$\hat Q^{(i)} = \sum\limits_{p=1}^{k} 2^{p-1} \hat \sigma_z^{(p,j)} -  \hat I /2$. 

In non-adiabatic quantum annealing~\cite{David21}, the binary encoding offers the so-called ``Goldilocks zone" in the numerical simulation, where the finding probability of the first excited state is superior to those of the ground state and the second excited state, which cannot be seen in the Hamming-weight encoding. This is due to the smaller energy gap of the binary encoding than the Hamming-weight encoding. 
However, on the D-Wave hardware, Hamming-weight encoding provides significantly shorter vectors than the binary encoding~\cite{David21}. 

In this paper, we offer an alternative encoding, ``one-hot encoding", for the shortest vector problem. 
If we limit the range of an integer coefficient $x_i$ of each basis vector as $-k \leq x_i \leq k$, then $2k+1$ qubits are required for each basis, resulting in
a total of $(2k+1)n$ qubits for the $n$-dimensional lattice problem. 
The diagonal operator $\hat{Q}^{(i)}$ is defined as 
\begin{align}
\hat{Q}^{(i)} = \sum_{p=0}^{2k}(-k+p)\hat q^{(p,i)}. 
\end{align}
In the one-hot encoding, a state of $2k + 1$ qubits represents an integer $x_i \in [-k, k]$, which can be given by $x_j = (-k + p)$, where an eigenvalue of the $p$-th operator $\hat q^{(p,i)}$ is $+1$ and eigenvalues of all the other operators are $0$. 

In contrast to the binary encoding and the Hamming-weight encoding~\cite{David21,Ura22}, the one-hot encoding needs the penalty term, which enforces the condition where only one operator $\hat q^{(p,i)}$ has the eigenvalue $+1$ and the others are zeros.  
This constraint could be included as a penalty Hamiltonian 
\begin{align}
 \hat H_{\text{penalty}} =  \lambda \sum_{i=1}^{n} \hat P_i, 
\end{align}
where the penalty of the one-hot constraint is given by 
\begin{align}
  \hat P_i \equiv \left( \sum_{p=0}^{2k} \hat q^{(p,i)} - 1 \right) ^2 , 
\end{align}
which can be expanded in the Pauli-$Z$ operator as 
$\hat P_i  = \left[ \sum_{p=0}^{2k} \hat \sigma_z^{(p,i)}  + (2k - 1) \right] ^2 /4$. 
If a configuration of spins violates the constraint, a penalty term increases the energy, where the strength of the penalty is controlled by the coefficient $\lambda (> 0)$.

The problem Hamiltonian for the SVP in the one-hot encoding is then given by 
\begin{align}
  \hat{H}_{\rm SVP}  = & \hat H_{\rm p} + \hat  H_{\text{penalty}} 
  \nonumber
  \\ 
  = & J \sum_{i,j=1}^n G_{ij} \hat{Q}^{(i)}\hat{Q}^{(j)}  +  \lambda \sum_{i=1}^{n} \hat P_i. 
  \label{HSVP}
\end{align}
Since the SVP is mapped to the first excited state search problem, we need to care the strength of the penalty coefficient $\lambda$. 
Suppose that two operators of $\hat q^{(p,i)}$ has the eigenvalue $+1$ in $\hat  H_{\text{penalty}}$ in a certain dimension, but the other dimension satisfies the constraint. 
In this case, the penalty value is $\lambda$, which is the eigenvalue of the first excited state of $\hat  H_{\text{penalty}}$. 
In the context of the first excited state search in the SVP, this strength $\lambda$ should be at least larger than the first excited state energy $E_{\rm p}^{(1)}$ in $\hat H_p$, i.e, $\lambda > E_{\rm p}^{(1)}$. 
If this is the case, the first excited state of the total Hamiltonian $\hat H_{\rm SVP}$ is the solution of the SVP. 
Even if we do not know the value of $E_{\rm p}^{(1)}$, 
we can determine $\lambda$ from the size of the search area of the lattice point we are considering. 

\section{Folded-spectrum method}

\begin{figure}[tbp]
  \includegraphics[keepaspectratio, width=80mm]{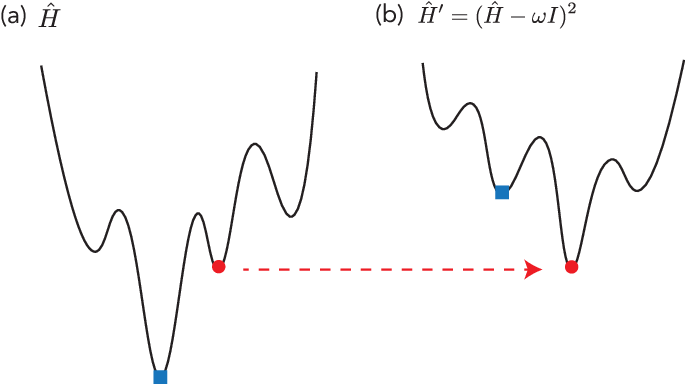}
\caption{
Schematics of the folded-spectrum (FS) method. The left and right figures show the energy landscapes of the original Hamiltonian $\hat H$ and the FS operator $\hat H' \equiv (\hat H - \omega)^{2}$, respectively. 
The eigenstate of $\hat H$, the energy of which is closest to $\omega$, is a target we are searching (red circle). (Blue square is the ground state energy in the original Hamiltonian.)
The energy landscape is folded at $\omega$ in the FS operator. 
If we prepare an appropriate value of $\omega$, the target state becomes the ground state in the FS operator. 
Schematics here are just conceptual.
}
\label{FSMethod}
\end{figure}

The folded-spectrum (FS) method is a useful method to search a state around a desired energy, which is investigated in the field of quantum chemistry~\cite{Wang1994}. 
This method employs the FS operator 
\begin{align}
  \hat{H}^\prime = (\hat{H} - \omega \hat I)^{2m}, 
  \label{FSHamiltonian}
\end{align}
to search the spectrum in $\hat H$ around the parameter $\omega$, where $m$ is the positive integer. 
Figure~\ref{FSMethod} illustrates the concept of the FS method. 
Consider the problem to search an excited state around the target energy $\omega$ in the original Hamiltonian $\hat H$, where $\hat H  |\phi_i\rangle = E_i | \phi_i\rangle$ holds. 
The FS operator $\hat H^\prime$ shares the eigenstates $|\phi_i \rangle$ in the target Hamiltonian $\hat H$, the eigenvalue of which is given by $(E_i - \omega)^2$, where the energy spectrum is folded at $\omega$. 
The ground state of the FS operator is then the eigenstate of $\hat H$, which is the closest to $\omega$. 
Therefore, by introducing the FS operator, the excited state search of $\hat H$ around $\omega$ is mapped to the ground state search of $\hat H'$. 

The quantum imaginary time evolution (QITE) method is a well-known technique for creating a Gibbs state and extracting a ground state, which can be implemented in NISQ algorithms~\cite{Yuan2019theoryofvariational, McArdle2019,Motta2020, PRXQuantum.2.010317, Motta2020,PhysRevA.104.032413}.
The time evolution of quantum systems is described by the Schr\"odinger equation 
\begin{align}
  i \dfrac{\partial \ket{\psi(t)}}
  {\partial t} = \hat{H} \ket{\psi(t)}, 
\end{align} 
the solution of which for the time-independent Hamiltonian $\hat H$ is given by 
$  \ket{\psi(t)} =  \sum_{i} c_i e^{-i E_i t }\ket{\phi_i} $,
where $E_i$ and $\ket{\phi_i}$ are the $i$-th eigenvalue and eigenstate of $\hat{H}$ and we take the unit $\hbar = 1$. 
By using the imaginary time $\tau=it$, the solution can be represented as 
\begin{align}
  \ket{\psi(\tau)} =
  \sum_{i} c_i e^{-E_i \tau / \hbar} \ket{\phi_i}. 
\end{align} 
Since the contributions of all the excited states decay exponentially in a quantum state in the QITE, 
we can obtain the ground state after the long imaginary-time evolution, as 
\begin{align}
  \dfrac{\ket{\psi(\tau)}}
  {\sqrt{\braket{\psi(\tau)|\psi(\tau)}}}
  \xrightarrow[\tau\to\infty]{}
  \ket{\phi_0} .
\end{align}
As $\tau$ approaches infinity, any initial state with a non-zero overlap with the ground state converges
towards the ground state of the Hamiltonian. 

The QITE is useful not only for the ground state search but also for the excited state search.  
Indeed, by utilizing this FS operator $\hat{H}^\prime$ to evolve the quantum state with the imaginary
time, the state converges to the ground state of the FS operator $\hat{H}^\prime$, which is the excited state of the original Hamiltonian $\hat H$ around $\omega$. 
The FS method with QITE in the NISQ algorithm has been discussed in the field of the quantum chemistry~\cite{Tsuchimochi23}. 
For the NISQ algorithm, the excited state search is also discussed in the subspace-search variational quantum eigensolver (VQE)~\cite{Nakanishi2019} and a folded-spectrum VQE~\cite{Tazi2024}. 

\section{Search and Bound}

The solution to the SVP corresponds to a non-trivial first excited state of the SVP Hamiltonian. 
Throughout this study, we propose a quantum method for searching the solution of the SVP with the FS method. 
For the FS method, finding a relevant value of the parameter $\omega$, which should be close to the unknown value of the norm of the shortest vector, is vital. 
We here propose a method, search and bound, to find an appropriate value of the parameter $\omega$. 
In the following, the lower bound and upper bound for the search and bound are taken as two real numbers $\alpha$ and $\beta$ with the relation $\alpha < \beta$, respectively. 

The energy of the Hamiltonian for the lattice problem corresponds to the square of the norm of the lattice vectors. 
Since the ground state of the Hamiltonian $\hat H$ is a trivial zero vector, the ground state eigenvalue is $E_0 = 0$. 
For finding the parameter $\omega$,  
the lower bound $\alpha$ should be then fixed as a sufficiently small positive real number $\varepsilon$, such as the machine epsilon, which is sufficient to satisfy the relation $E_1 > \alpha ( > 0)$ in numerical calculations. 

\begin{figure}[tbp]
  \includegraphics[keepaspectratio, width=90mm]{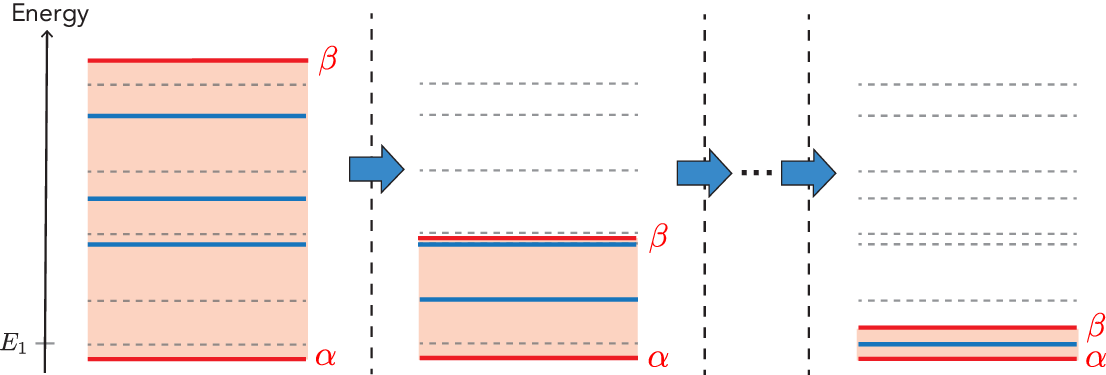}
\caption{Schematics of the search and bound for the SVP, where the range from the lower bound $\alpha$ to the upper bound $\beta$ is narrowed down to an appropriate range to extract the first excited-state energy $E_1$ alone. 
In the SVP, a trivial ground state gives the zero vector, and the lower bound $\alpha$ should be fixed such as a machine epsilon.
The upper bound $\beta$ is updated by using the search and bound. 
In the search and bound operations, a part of the eigenvalues in the SVP Hamiltonian $\hat H_{\rm SVP}$ can be detected (thick blue lines), where 
thin dashed horizontal lines represent the energy spectra in $\hat H_{\rm SVP}$. 
}
\end{figure}

If we find an appropriate upper bound $\beta$ such that the following relation holds 
\begin{align}
  \alpha < E_1 < \beta < E_2 < \cdots ,
\end{align}
the FS method with the parameter $\omega \in [ \alpha, \beta]$ will provide the first excited state with the energy $E_1$, where $E_2$ is the second excited state energy. 
Initially, $\beta$ should be taken as a value sufficiently larger than the unknown first-excited-state energy. 
The upper bound $\beta$ is then updated by the following search and bound operations. 

Suppose that the $m$-eigenvalues of the Hamiltonian $\hat H$ are in the range of $\alpha$ and $\beta$, such as $\alpha < E_{i_1,\cdots,i_m} < \beta$. 
If we take an appropriate value of the parameter $\omega$ as $\omega \in [\alpha, \beta]$, 
the FS method will provide one of the eigenstates with the energy $E_i \in \{ E_{i_1}, E_{i_2},\cdots,E_{i_m} \}$, the state of which depends on the parameter $\omega$. 

The search operation is to find an appropriate upper bound $\beta$. 
We divide the region  $[\alpha, \beta]$ into small sections, where we suppose that $n_\omega$-parameters are prepared such that $\omega_{1, \cdots, n_\omega} \in [\alpha, \beta]$. 
After executing the FS method with each $\omega_{1, \cdots, n_\omega}$, 
we will obtain a set of eigenvalues $\{E_{i_j^\prime} \}_{j=1}^{m^\prime}$ that satisfy
$\alpha < E_{i_1^\prime,\cdots,i_{m^\prime}^\prime} < \beta$. 
Note that the set obtained by this method may not cover all the eigenvalues in $[\alpha, \beta]$ but a subset of $\{E_{i_j}\}_{j=1}^m$, i.e., $\{E_{i_j^\prime} \}_{j=1}^{m^\prime} \subseteq \{E_{i_j}\}_{j=1}^m$ with $m^\prime \leq m$. 
The search operation is to extract the minimum value $E_{i_1^\prime}$ among the set $\{E_{i_j^\prime} \}_{j=1}^{m^\prime}$. 

The bound operation is to update the upper bound $\beta$. 
We update the upper bound as $\beta \leftarrow E_{i_1^\prime} + x$ with a small positive real number $x$. 
Here, $E_{i_1^\prime}$ is the lowest eigenvalue found in the search operation, and $x$ is a buffer to ensure that the first excited state energy $E_1$ can be sufficiently in the range $[\alpha, \beta = E_{i_1^\prime} + x]$, even in the case for $E_{i_1^\prime} = E_{1}$, to exclude the possibility where $E_{1}$ is missed. 
This process is repeated until only one eigenvalue ($m^\prime=1$) is obtained with the FS method ($m^\prime=1$). 
The obtained value is then $E_1$, and the obtained state is the first-excited state. 

In the implementation, we set a maximum number of the iteration, 
and if $m^\prime=1$ is reached during the iteration, we find the solution of the SVP. 
However, in practice, there may be cases where $m^\prime=1$ is not achieved even after reaching the maximum iteration limit. 
In this case, we employ the minimum value of the obtained eigenvalues $\{E_{i_j}^\prime\}_{j=1}^{m^\prime}$ as an approximate value for the first excited energy and perform the FS method with this value as the upper bound again. 
During the interation, we can reduce the number of parameters $\omega_{1, \cdots, n_\omega}$ to search the eigenstates, because the number of the eigenstates in the region $[\alpha, \beta]$ are narrowed down through the repetation of the search and bound operations.

 \section{Quantum Imaginary-time Evolution for SVP} 

 \begin{figure}[tbp]
   \centering
   \includegraphics[keepaspectratio, width=90mm]{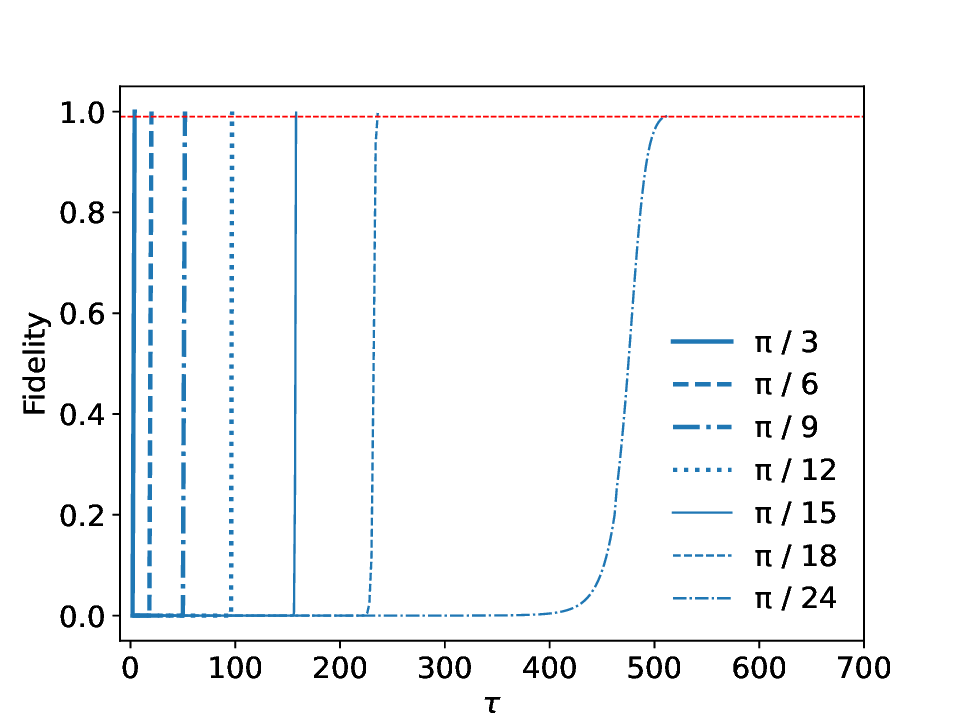}
   \caption{Imaginary-time dependence of the fidelity with respect to the shortest vector state. 
   We used $m=1$ for the FS operator $\hat H^\prime = (\hat H - \omega)^{2m}$, and $\lambda = 2.5$ for the penalty term. 
   In the SVP, we take $(n,k) = (2,2)$. 
   The dotted horizontal line represents a threshold $f_{\rm th} = 0.99$ we set, where 
   we applied the imaginary time evolution until the fidelity exceeds the threshold. 
   }
   \label{time_vs_fidelity}
 \end{figure}

\begin{figure}[tbp]
  \centering
   \includegraphics[keepaspectratio, width=90mm]{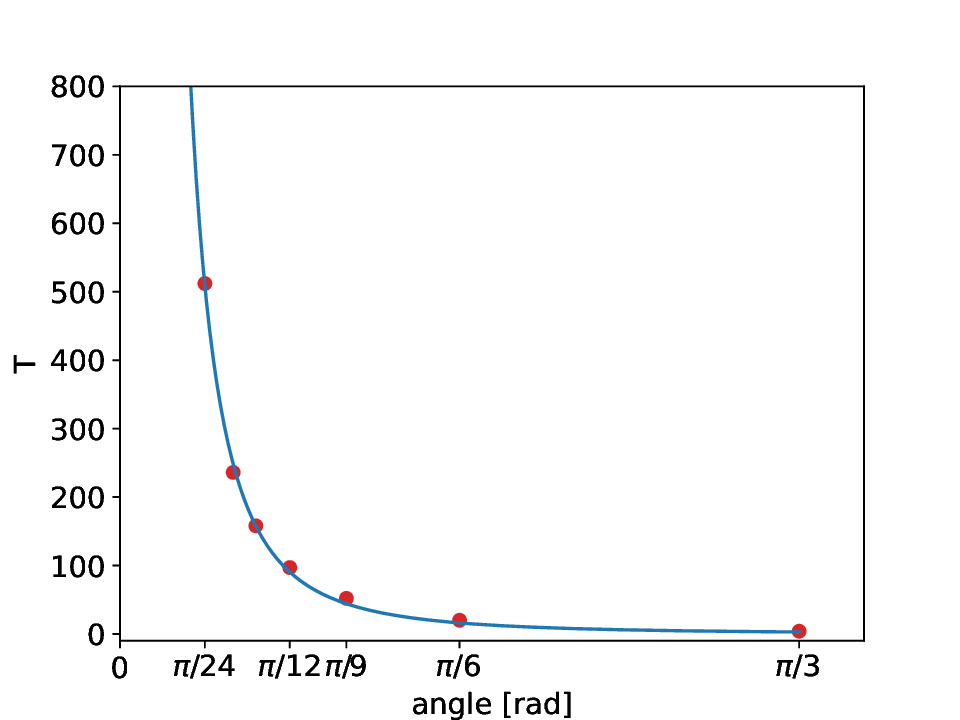}
  \caption{Angle dependece of the evolution time $T$ such that the fidelity exceeds a threshold $f_{\rm th} = 0.99$. The time $T$ shows the power-law with respect to the angle $\theta$. }
  \label{angle_vs_T}
\end{figure}
 
 We present numerical results of the FS method of QITE for a two-dimensional SVP ($n=2$). 
 The parameter $k$ denoting the range of coefficients of the basis vectors is chosen as $2$, where $10$-qubits are prepared. 
 The two basis vectors, $\bm{b}_1$ and $\bm{b}_2$, are normalized to $\|\bm{b}_{1,2}\|=1$, the angle between which is characterized by $\theta$. 
 The penalty strength $\lambda$ is chosen such that the minimum penalty energy of a state that violates a constraint, the Hamming-weight of which is $2$, is larger than  
 the maximum value of the square of the norm of the lattice vector in the area we are searching.  
 which depends on $k$ and $\theta$.

 The imaginary-time dependence of the fidelity of the shortest vector state are shown in Fig.~\ref{time_vs_fidelity}, the simulation of which is executed by solving the imaginary time Schr\"odinger equation. 
 In general, in good and bad bases, the SVP is known to be relatively easy and difficult to solve, respectively. 
 In the quantum approach, as the angle between lattice bases decreases, the energy level spacing in the SVP Hamiltonian are narrower, including the energy gap between the ground state energy $E_0$ and the 1st-excited state energy $E_1$. 
 Consequently, in a good basis case with a large angle such as $\theta = \pi/3$, the fidelity of the shortest vector state, i.e., first-excited state, is rapidly increased to the threshold $f_{\rm th} = 0.99$ that we set,  
 while a longer evolution time is required for the fidelity exceeding the threshold $f_{\rm th}$  in a bad basis case with a small angle such as $\theta = \pi/24$. 
The evolution time $T$ that is necessary to exceed the fidelity threshold $f_{\rm th}$ becomes longer with decreasing the angle $\theta$ according to the power law scaling: $T = a \theta^{-b}$ with $a = 2.98$ and $b = 2.41$ (FIG.\ref{angle_vs_T}).  
 The reason why the angle we discuss is in $(0,{\pi}/{3}]$ is summarized in the Appendix. 
 Since we employ the folded-spectrum method for the quantum imaginary time evolution with a square form of the Hamiltonian, $(\hat{H}-\omega I )^2$, 
 the energy level spacing less than $1$ is compressed compared with the bare SVP Hamiltonian, which requires the longer evolution time in the FS method. 
 
\begin{figure}[tbp]
   \includegraphics[keepaspectratio, width=90mm]{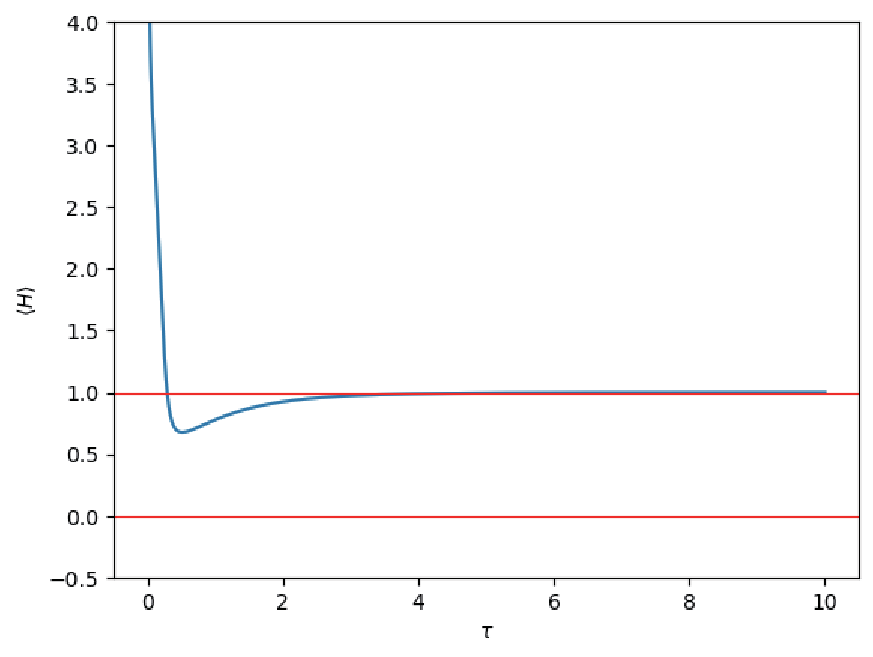}
   \caption{Imaginary-time dependence of the expectation value of the Hamiltonian, which corresponds to the square of the norm of the lattice vector, by using the folded-spectrum method. Here, we used the efficient SU2 circuit for the variational quantum imaginary-time evolution.}
   \label{fs_energy_image_v2}
 \end{figure}

The imaginary-time evolution algorithm for the SVP can be implemented to the quantum computers with NISQ algorithms. 
The key idea is to reproduce an operator of the imaginary time evolution, which is a non-unitary operator, with unitary operators developed by quantum gates~\cite{Yuan2019theoryofvariational, McArdle2019,Motta2020, PRXQuantum.2.010317, Motta2020}. 
By using the variational quantum techniques with an ansatz circuit for the NISQ algorithm, 
we can successfully find a shortest vector as the first excited state of the SVP Hamiltonian (Fig.~\ref{fs_energy_image_v2}). 
Here, we used the VarQITE in qiskit, which is a time evolution class that performs variational calculations based on McLachlan's variational principle. 
In this simulation, an efficient SU2 circuit is employed as an ansatz circuit (FIG.\ref{Ansatz_circuit}), where we used the number of the repetition as $2$.

 \begin{figure}[tbp]
   \includegraphics[keepaspectratio, width=90mm]{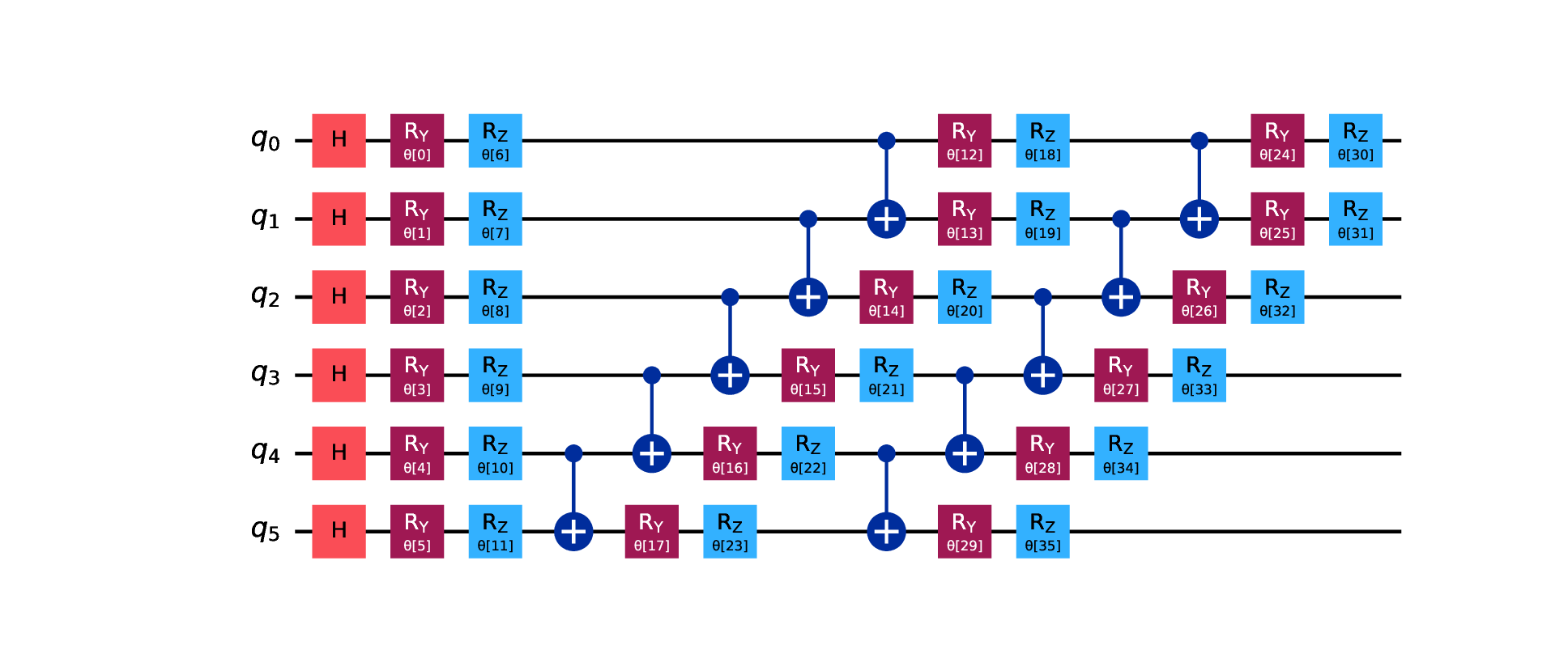}
\caption{Ansatz circuit, the efficient SU2, used in the simulation shown in Fig.~\ref{fs_energy_image_v2}. } 
   \label{Ansatz_circuit}
 \end{figure}

 \section{Application to Quantum Annealing and Simulated Annealing}

In the NISQ algorithms for the SVP, we could employ not only the QITE but also the subspace-search VQE~\cite{Nakanishi2019} and a folded-spectrum VQE~\cite{Tazi2024}, which are excited state search algorithms of the Hamiltonian. 
We here propose and demonstrate alternative implementations of the FS method for the SVP, in particular, the quantum annealing and simulated annealing, other than the NISQ algorithms.

In the quantum annleaing, we employ the following time-dependent Hamiltonian~\cite{PhysRevE.58.5355} 
\begin{align}
\hat H(x) = f_{\rm p} (t) \hat H_{\rm SVP} + f_{\rm d} (t) \hat H_{\rm d}, 
\label{QAHamiltonian}
\end{align}
where $\hat H_{\rm d} \equiv - h_x \sum\limits_{p,i} \hat \sigma_x^{(p,i)}$ is the driver Hamiltonian and the boundary conditions of the scheduling function $f_{\rm p,d}$ are given by 
\begin{align}
f_{\rm p} (t) = \begin{cases} 0 & (t = 0) , \\ 1 & (t=T), \end{cases}  
\quad 
f_{\rm d} (t) = \begin{cases} 1 & (t = 0), \\ 0 & (t=T) , \end{cases} 
\end{align}
with the annealing time $T$. 
For example, a typical scheduling function is given as $f_{\rm p} (t) = t/T$ and $f_{\rm d} (t) = 1-t/T$. 
In the conventional quantum annelaing, the initial state is chosen as a trivial ground state of the driver Hamiltonian $\hat H_{\rm d}$, which is given by $|+\rangle^{\otimes n}$ with $|+\rangle = (|0\rangle +|1\rangle )/\sqrt{2}$. 
If the annealing time $T$ is taken to satisfy the adiabatic condition~\cite{BAPST2013127}, 
we can find the ground state of the problem Hamiltonian. 

In the so-called ``not-so-adiabatic quantum computation", or non-adiabatic quantum annealing, for the first excited state search~\cite{Joseph20,David21}, 
the adiabatic condition is violated on purpose for the sweeping, where the Landau-Zener tunneling from the ground state to excited states are dared to cause. 
The time sweep optimization 
to create a ``Goldilocks" zone, where the probability to find the first excited state is larger than those of the ground state and other excited states, are needed~\cite{Joseph20,David21}. However, the existence of the ``Goldilocks" zone depends on the encoding~\cite{David21}, and the ideal zone obtained in the numerical simulation is hard to reproduce in the realistic quanum aneealing machine, such as the D-wave 2000Q quantum annealer~\cite{8728085,David21}. 

In the excited state search for the SVP with the quantum annealing~\cite{Ura22}, the first excited state is prepared as an initial state. 
In this case, satisfying the adiabatic condition and avoiding the decay from the first excited state to the ground state are needed, where 
the quantum annealing with the capacitively-shunted flux qubits will be helpful for a long coherent time~\cite{Matsuzaki_2020}. 

\begin{figure}[tbp]
\includegraphics[keepaspectratio, width=90mm]{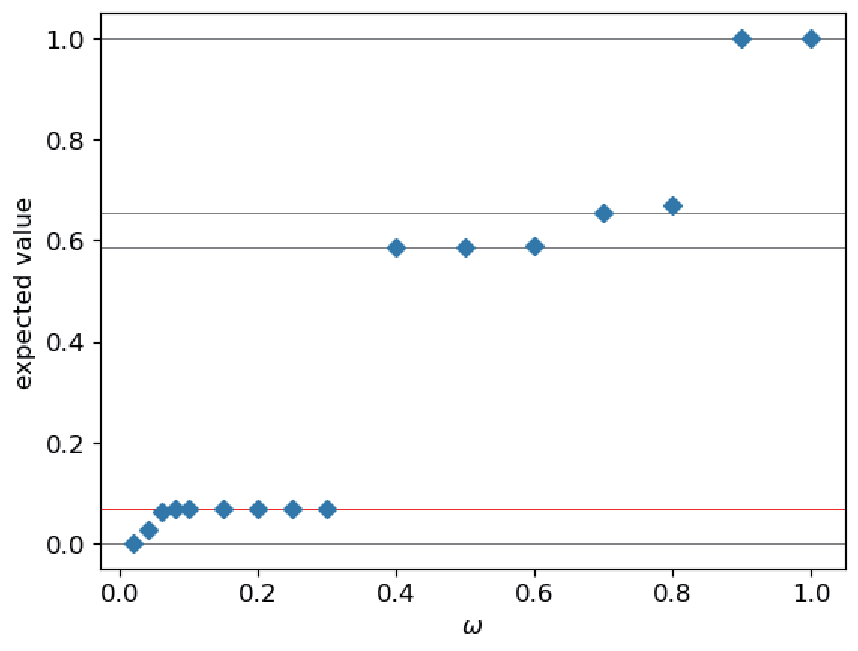}
\caption{The $\omega$-dependece of the expectation value of the SVP Hamiltonian obtained from the FS quantum annealing. Horizontal lines represent the eigenvalues of the SVP Hamiltonian, the lowest and second lowest of which are the ground state energy and the first excited state energy that corresponds to the shortest vector, respectively.}
\label{qa_omega_ev}
\end{figure}

In the quantum annealing with the FS method, we employ the conventional quantum annealing for the ground state search, where the adiabatic conditions is important in principle. 
On the other hand, the problem Hamiltonian $\hat H_{\rm SVP}$ in \eqref{QAHamiltonian}, which is the so-called quadratic unconstrained binary optimization (QUBO) formulation, is replaced with $\hat H'$ in \eqref{FSHamiltonian}, which is the higher-order binary optimization (HOBO) formulation. 
Using the quantum annealing simulation with the FS method, we can estimate the shortest vector, which can be seen in the first plateau with respect to the parameter $\omega$ (Fig.~\ref{qa_omega_ev}). 

This FS method for the SVP is also applied to the simulated annealing (SA). 
With the slower cooling rate, the frequency is high enough to find the shortest vector. 
While the approach is different from the non-adiabatic quantum annealing, where the shortest vector was not detected in the D-Wave 2000Q quantum annealer~\cite{8728085,David21}, our SA method has the histgram at the shortest vector. 

Table~\ref{table1} summarizes the study on the SVP with the quantum computations. 
While the non-adiabatic quantum annealing is applied to the qudit, Hamming-weight, as well as binary encodings~\cite{Joseph20,David21}, 
the excited-state search quantum annealing is implemented in the Hamming-weight encoding~\cite{Ura22}, 
and the FS method with the QITE, QA and SA is proposed with the one-hot encoding in this study. 
The performance of the quantum algorithm for the SVP will depend on the encoding, since the encoding affects the energy spectrum and energy gap in the SVP Hamitlonian. 
This was highlighted in the study of the non-adiabatic quantum annelaing in the Hamming-weight encoding and the binary encoding~\cite{David21}. 
As in the table~\ref{table1}, the character of the encoding in the SVP Hamiltonian and the performance of the quantum algorithms have not been systematically investigated and there is room for further study. 
Furthermore, fault-tolerant quantum algorithms, such as the quantum phase estimation~\cite{1995quant.ph.11026K} will be helpful to find the shortest vector in future, since the SVP is already embedded in the problem Hamiltonian.

\begin{figure}[tbp]
\includegraphics[keepaspectratio, width=90mm]{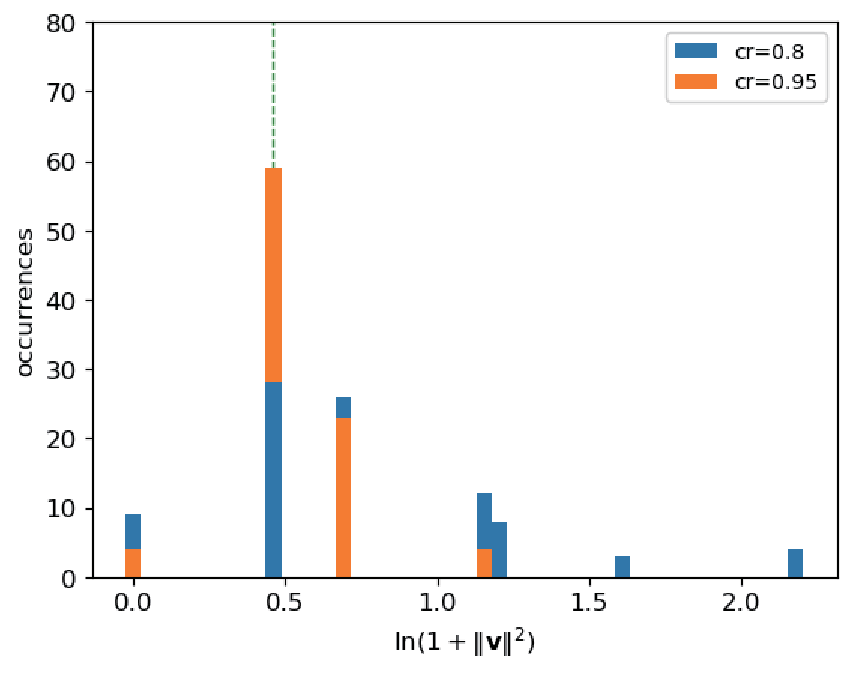}
\caption{Histgram obtained from the FS simulated annealing (SA), where $||{\bf v}||^2$ is the square of the norm of the lattice vector generated by the SA. We employed the cooling rates $0.8$ and $0.95$. The dashed vertical line represents the position of the target shortest vector. 
}
\label{Ansatz_circuit}
\end{figure}

\begin{table*}[tbp]
    \centering
    \begin{tabular}{rcccccc} \hline\hline 
    \,\, encoding  & \,\, non-adiabatic QA  \,\, & \,\, excited-state QA \,\, & FS+(QITE/QA/SA) 
    \\ \hline 
     qudit & $\checkmark$~\cite{Joseph20} & & & & 
    \\
    Hamming-weight & $\checkmark$~\cite{David21}  & $\checkmark$~\cite{Ura22}  & 
    \\
    binary & $\checkmark$~\cite{David21}  & & 
    \\ 
    one-hot  & & &  $\checkmark$ (this study) &   \\ 
               \hline\hline
    \end{tabular}
\caption{Study of the shortest vector problem with the quantum calculation. 
    Non-adiabatic quantum annealing (QA) is to utilize the fast annealing to generate the excited states from the ground state~\cite{Joseph20,David21}. The excited state QA is to used the adiabatic annealing to search the excited state in the problem Hamiltonian from a corresponding excited state in the driver Hamiltonian~\cite{Ura22}. 
In the present study, we employ the folded spectrum method with the quantum imaginary time evolution (QITE), quantum annealing (QA), and simulated annealing (SA), where the ground state search algorithms are used in the folded-spectrum Hamiltonian. 
    Several encodings to the Hamiltonian can be proposed, such as qudit, Hamming-weight, binary and one-hot. 
    } 
    \label{table1}
\end{table*}

\section{conclusion}
To implement the shortest vector problem (SVP) to quantum computers as the first-excited state search problem, the non-adiabatic quantum annealing (QA) or the first-excited state QA starting from the first excited state in the driver Hamiltonian have been discussed. 
We here proposed an alternative approach to solve the SVP by leveraging the quantum imaginary time evolution method. 
We extended the folded-spectrum (FS) method, which is widely discussed in the fields of the quantum chemistry, to solve the SVP. 
We also proposed the search and bound to find an appropriate parameter in the FS method for the SVP. 
We demonstrated that the quantum imaginary-time evolution with the FS method can be applicable to the SVP by using the simulation of the imaginary-time Schr\"odinger dynamics as well as the variational quantum algorithm. 
This FS method can be also applicable to the SVP with the use of the quantum annealing and the simulated annealing. 
These results highlight their potential applicability in quantum computing frameworks. 

We also proposed an alternative encoding, the one-hot encoding, to the SVP Hamiltonian. 
While the encoding is important for the quantum algorithms in the SVP, because the energy spectrum directly affects on the performance of quantum algorithms, there is room for further study to charaterize the performance of the quantum algorithms with various encodings.

\begin{acknowledgments} 
This work was supported by JST, PRESTO Grant Number JPMJPR211A, Japan.
\end{acknowledgments}

\appendix

\section{Angle between two basis vectors in the two-dimensional lattice}

In this appendix, we discuss the relationship between the value of the first excited state energy $E_1$ and the angle $\theta$ between two basis vectors $\bm{b}_{1,2}$ in the two-dimensional lattice case. 
We assume that the norm of $\bm{b}_{1,2}$ is normalized as $\|\bm{b}_{1,2}\|=1$. 
 
 First, for $0 < \theta \leq \pi/3$, the first excited
 state, which is the solution of the SVP, is the lattice point vector represented
 by $\pm (\bm{b}_1-\bm{b}_2)$. The first excited state energy $E_1$ of the SVP Hamiltonian corresponds to $\| \bm{b}_1 - \bm{b}_2 \|^2$, which is expressed as $E_1 = 2 - 2 \cos \theta$.
 
For $\pi/3 < \theta \leq 2\pi/3$, the norm of the lattice point vectors created by $\bm{b}_{1,2}$ is greater than $1$. 
As a result, the first excited state energy is $E_1 = 1$, the state of which corresponds to the basis vectors $\pm \bm{b}_{1,2}$. 
 
 Finally, for $2\pi/3 < \theta \leq \pi$, the first excited
 state is the lattice point vector represented
 by $\pm (\bm{b}_1+\bm{b}_2)$. The square of the norm of this vector is $E_1 = \| \bm{b}_1 + \bm{b}_2 \|^2 =  2 + 2 \cos \theta$. 

 The angle dependence of the first excited state energy is symmetric, 
 and its dependence is absent in $\pi/3 < \theta \leq 2\pi/3$. 
 This is the reason whey we chose the angle $0 < \theta \leq \pi/3$ in Figs.~\ref{time_vs_fidelity} and~\ref{angle_vs_T}.

 At $\theta = \pi/3$, 
 the norm of the lattice point vector formed by the sum of two basis vectors equals $1$, 
 leading to the first excited state being the $6$-fold degenerate. 
Consequently, at angles approximately around $\pi/3$, the energy spectra of the first excited state and the other excited states become closer. 
However, in the search and bound with the FS method for the QITE in $(n,k) = (2,2)$, 
we can efficiently find an appropriate value of $\omega$ and the first excited state. 
This is in contrast to the case with $\theta \simeq 0$, where the spectra of the ground state, the first-excited state and the other state become close, and the long imaginary-time evolution should be necessary.

\bibliography{ref}

\end{document}